\documentclass[12pt]{article}

\usepackage{cite}

\usepackage{times}

\usepackage{amsmath}
\usepackage{multicol}
\usepackage{multirow}
\title{GPU Accelerated Keccak (SHA3) Algorithm}

\author{
	{\footnotesize Canhui Wang $^{\dagger}$, Xiaowen Chu $^{\dagger,\star}$} \\
	{\footnotesize \textit{$^{\dagger}$ Computer Science, Hong Kong Baptist University, Hongkong, China}}\\
	{\footnotesize \textit{$^{\star}$ HKBU Institute of Research and Continuing Education, Shenzhen, China}}\\
	{\footnotesize \textit{\{chwang, chxw\}@comp.hkbu.edu.hk}}
}

\date{}

\begin{document} 

\baselineskip13pt
\maketitle

\begin{abstract}
	
	Hash functions like SHA-1 or MD5 are one of the most important cryptographic primitives,
	especially in the field of information integrity. Considering the fact that increasing methods have been proposed to break these hash algorithms, a competition for a new family of hash functions was held by the US National Institute of Standards and Technology. Keccak was the winner and selected to be the next generation of hash function standard, named SHA-3.
	
	We aim to implement and optimize Batch mode based Keccak algorithms on
	NVIDIA GPU platform. Our work consider the case of processing multiple hash tasks at once and implement the case on CPU and GPU respectively. Our experimental results show that GPU performance is significantly higher than CPU is the case of processing large batches of small hash tasks.

\end{abstract}

\section{Introduction}

Security techniques \cite{saeed2018examine, kumar2018data} have been acknowledged to be an integral part in many fields i.e., business, national defense, military and etc. One special part of cryptographic algorithms is the hashing family that is important, especially in the field of modern information security where have a wide range of applications, i.e., digital signatures, message authentication codes, password authentication and etc. Unfortunately, today increasing existing hash algorithms like MD5, SHA-1 and so on are at high risk of being cracked. To improve the
security of the hash algorithms, a new SHA-3 \cite{dworkin2015sha} algorithm driven from KECCAK has been proposed to replace the older hash functions. Hash functions are unique in the way that an output is generated. A message is broken down into a number of blocks and the hash function consumes each block of the message into some type of internal state, with a final output produced after the last block is consumed. This structure is difficult to parallelize. In this case, the inputs could range from a few bytes to a few terabytes, and using a sequential hash function is not the best choice. A function with a tree hashing mode could be used to significantly reduce the amount of time that is required to compute the hash.

There are classes of problems that may be expressed as data-parallel computations with high arithmetic intensity where a CPU is not particularly efficient. Multi-core CPUs excel at managing multiple discrete tasks and processing data sequentially, by using loops to handle each element. Instead, the architecture of GPU maps the data to thousands of parallel threads, each handling one element. This architecture looks ideal for our fast algorithm implementation.

\section{Related Work}

 Lowden \cite{lowden2015design} focus on the exploration and analysis of the Keccak tree hashing mode on a GPU platform. Based on the implementation, there are core features of the GPU that could be used to accelerate the time it takes to complete a hash due to the massively parallel architecture of the device. In addition to analyzing the speed of the algorithm, the underlying hardware is profiled to identify the bottlenecks that limited the hash speed. The results of their work show that tree hashing can hash data at rates of up to 3 GB/s for the fixed size tree mode.

Qinjian et al., \cite{li2012implementation} propose a GPU based AES implementation. In their implementation, the frequently accessed T-boxes were allocated on on-chip shared memory and the granularity that one thread handles a 16 Bytes AES block was adopted. Finally, they achieve a performance of around 60 Gbps throughput on NVIDIA Tesla C2050 GPU, which runs up to 50 times faster than a sequential implementation based on Intel Core i7-920 2.66GHz CPU.

Kaiyong et al., \cite{zhao2014g} develop G-BLASTN, a GPU-accelerated nucleotide alignment tool based on the widely used NCBI-BLAST. G-BLASTN can produce exactly the same results as NCBI-BLAST, and it has very similar user commands. Compared with the sequential NCBI-BLAST, G-BLASTN can achieve an overall speedup of 14.80X under ‘megablast’ mode. They \cite{zhao2010gpump} also propose to exploit the computing power of Graphic Processing Units (GPUs) for homomorphic hashing. Specifically, they demonstrate how to use NVIDIA GPUs and the Computer Unified Device Architecture (CUDA) programming model to achieve 38 times of speedup over the CPU counterpart. They also develop a multi-precision modular arithmetic library on CUDA platform, which is not only key to our specific application, but also very useful for a large number of cryptographic applications.

Xinxin et al., \cite{mei2014benchmarking, mei2017dissecting} propose a novel fine-grained benchmarking approach and apply it on two popular GPUs, namely Fermi and Kepler, to expose the previously unknown characteristics of their memory hierarchies. They also investigate the impact of bank conflict on shared memory ac- cess latency.

Thuong et al., \cite{yang2017parallel} implements a high speed hash function Keccak
(SHA3-512) using the integrated development environment CUDA for GPU is proposed. In addition, the safety level of Keccak is also discussed at the point of Pre-Image Resistance especially. In order to implement a high speed hash function for password cracking, the special program is also developed for passwords up to 71 characters. Moreover, the throughput of 2-time hash is also evaluated in their work.

Chengjian et al., \cite{liu2018g} propose a graphics processing unit (GPU)-based implementation of erasure coding named G-CRS, which employs the Cauchy Reed-Solomon (CRS) code, to overcome the aforementioned bottleneck. To maximize the coding performance of G-CRS, they designed and implemented a set of optimization strategies, such as a compact structure to store the bitmatrix in GPU constant memory, efficient data access through shared memory, and decoding parallelism, to fully utilize the GPU resources.

Xiaowen et al., \cite{chu2009practical, chu2008practical} exploit the po- tential of the huge computing power of Graphic Processing Units (GPUs) to reduce the computational cost of network coding and homomorphic hashing. With their network coding and HHF implementation on GPU, they observed significant computational speedup in comparison with the best CPU implemen- tation. This implementation can lead to a practical solution for defending against the pollution attacks in distributed systems.

Cheong et al., \cite{cheong2015fast} contribute to the cryptography research community by presenting techniques to accelerate symmetric block ciphers (IDEA, Blowfish and Threefish) in NVIDIA GTX 690 with Kepler architecture. The results are benchmarked against implementation in OpenMP and existing GPU implementations in the literature. We are able to achieve encryption throughput of 90.3 Gbps, 50.82 Gbps and 83.71 Gbps for IDEA, Blowfish and Threefish respectively. Block ciphers can be used as pseudorandom number generator (PRNG) when it is operating under counter mode (CTR), but the speed is usually slower compare to other PRNG using lighter operations. Hence, they attempt to modify IDEA and Blowfish in order to achieve faster PRNG generation. The modified IDEA and Blowfish manage to pass all NIST Statistical
Test and TestU01 Small Crush except the more stringent tests in TestU01 (Crush and BigCrush).

\section{Preliminary}

The secure hash algorithm-3 (SHA-3) family is based on an instance of \textit{Keccak} algorithm that has been selected as the winner of the SHA-3 cryptographic hash algorithm competition by NIST in 2012. The  SHA-3 consists of four cryptographic hash functions, including SHA-3-224, SHA-3-256, SHA-3-384 and SHA-3-512, as well as two additional extendable output functions, SHAKE-128 and SHAKE-256. Specifically, the extendable output functions are different from hashing functions. It provides a flexible way to be adopt easily in according with the requirements of individual applications. In general, the hash functions play an important role in many fields, including digital signatures, pseudorandom bit generation etc.

\subsection{\textit{Keccak-p}}

The SHA-3 functions can be viewed as modes of $Keccak-p$ permutations, which are designed as the main components of various cryptographic functions. Two core parameters of $Keccak-p$ permutations are specified as $width$ and $round$. In this case, $width$ is denoted by $b$, meaning the fixed length of the permuted strings and $round$ is denoted by $n_r$, meaning that the number of iterations of an internal transformation.

The state of $Keccak-p[b,n_r]$ consists of $b$ bits. In addition, specifications in standard contain two other quantities related to $b$: $b/25$ and $log_2(b/25)$, denoted by $w$ and $l$, respectively. Seven possible cases for these variables that are defined for $Keccak-p[b,n_r]$ are given in Table 1.

\begin{table}[htbp]
	\centering
	\caption{The widths and other quantities of $Keccak-p[b,n_r]$}
	\label{my-label}
	\begin{tabular}{l|l|l|l|l|l|l|l}
		\hline
		b & 25 & 50 & 100 & 200 & 400 & 800 & 1600 \\ \hline
		w & 1  & 2  & 4   & 8   & 16  & 32  & 64   \\ \hline
		l & 0  & 1  & 2   & 3   & 4   & 5   & 6    \\ \hline
	\end{tabular}
\end{table}

It is convenient to represent the input and output states of the step mappings as five-by-five-$w$ array denoted as $A[x,y,z]$, meaning that an integer triple $(x,y,z)$ where $0 \leq x\leq 5$, $0 \leq y\leq 5$ and $0 \leq z\leq w$. A string can be denoted as $S$. An array is a representation of the string by a three-dimensional array and their relationship can be expressed in equation (1).

\begin{equation}
	A[x,y,z] = S[w(5y+x)+z]
\end{equation}

After representing a string into a state, next we need to operation the state. The specifications of operations, including $\vartheta$, $\rho $, $\pi  $, $\chi   $ and $\zeta   $ are discussed in the following section. Note that the algorithm for each step mapping takes a state array denoted by $A$. An return or output state array is denoted by $A'$. The size of the state is a parameter that is omitted from notation because $b$ is always specified when step mappings are invoked.

\noindent \textbf{Definition 1. $\vartheta $:} the input state array is denoted by $A$ and the output state array is denoted by $A'$. Then,

\begin{equation}
	C[x,z]=A[x,0,z] \oplus A[x,1,z] \oplus A[x,2,z] \oplus A[x,3,z] \oplus A[x,4,z]
\end{equation}

\begin{equation}
D[x,z]=C[(x-1)mod(5), z] \oplus C[(x+1)mod(5), (z-1)mod(w)]
\end{equation}

\begin{equation}
A'[x,y,z]=A[x,y,z] \oplus D[x,z]
\end{equation}

\noindent where $0 \leq x < 5$, $0 \leq y < 5$ and $0 \leq z < w$. The effect of the specification $\vartheta $ is to $XOR$ each bit in the state with parties of two columns in the array. In particular, for $A[x_0,y_0,z_0]$, the $x$-coordinate of one of the columns is $(x_0-1)mod(5)$ with the same $z$-coordinate $z_0$; while the $x$-coordinate of one of the columns is $(x_0+1)mod(5)$ with coordinate $(z_0-1)mod(w)$.

\noindent \textbf{Definition 2. $\rho  $:} the input state array is denoted as $A$ and the output state array is denoted by $A'$. $\forall t \in [0, 24)$, the specification of $\rho  $ is expressed as follows.

\begin{equation}
	A'[y,(2x+3y)mod(5), (z-(t+1)(t+2)/2)mod(w)]=A[x,y,z]
\end{equation} 

\noindent where $0 \leq z < w$ and the initial value $x=1$ and $y=0$. The effect of the specification of $\rho  $ is to rotate the bits of each lane by a length named $offset$, which depends on the fixed $x$ and $y$ coordinates of the lane. Equivalently, for each bit in the lane, the $z$ coordinate is modified by adding the $offset$ modulated by the lane size.

\noindent \textbf{Definition 3. $\pi  $:} the input state array is denoted by $A$ and the output state array is denoted by $A'$. The specification of $\pi$ is expressed as follows.

\begin{equation}
	A'[x,y,z]=A[(x+3y)mod(5),x,z]
\end{equation}

\noindent where $0 \leq x < 5$, $0 \leq y < 5$ and $0 \leq z < w$. The effect of the specification $\pi$ is to rearrange the positions of the lanes.

\noindent \textbf{Definition 4. $\chi   $:} the input state array is denoted by $A$, the output state array is denoted by $A'$. The specification of $\chi$ is expressed as follows.

\begin{equation}
	A'[x,y,z]=A[x,y,z] \oplus (A[(x+2)mod(5), y, z] \cdot (1 \oplus A[(x+1)mod(5), y, z]))
\end{equation}

\noindent where $0 \leq x < 5$, $o \leq y < 5$ and $0 \leq z < w$. Note that the dot in the equation (7) indicates integer multiplication which in this case is equivalent to the intended Boolean $AND$ operation. The effect of $\chi$ is to $XOR$ each bit with a non-linear function of two other bits in its row.

\noindent \textbf{Definition 5. $\zeta   $:} the input state array is denoted by $A$ and the output state array is denoted by $A'$. The specification of $\zeta$ is expressed as follows.

\begin{equation}
	RC[2^{j}-1] = rc(j+7i_r)
\end{equation}

\begin{equation}
	A'(0,0,z) = A(0,0,z) \oplus RC[z]
\end{equation}

\noindent where $0 \leq z < w$, $0 \leq j < l$. Note that within the specification of $\zeta$, a parameter determines $l+1$ bits of a lane value called the Round Constant and denoted by $RC$. Each of these $l+1$ bits is generated by a function that is based on a linear feedback shift register. The function is denoted by $rc$. The effect of $\zeta$ is to modify some of the bits of $A(0,0,z)$ where $0 \leq z < w$ in a manner that depends on the round index $i_r$.

Thus, given a state array $A$ and a round index $i_r$, the round function $Rnd$ is the transformation that results from applying the steps as follows.

\begin{equation}
	Rnd(A,i_r)=\zeta( \chi( \pi( \rho ( \vartheta (A) ) ) ), i_r )
\end{equation}

Note that the $Keccak-p[b, n_r]$ permutation consists of $n_r$ iterations of $Rnd$.

\subsection{Sponge}

The spong construction is a framework for specifying functions on binary data with arbitrary output length. The construction employs three components: an underlying function on fixed-length strings denoted by $f$, a parameter named the Rare denoted by $r$ and a padding function denoted by $pad$. These components form a sponge function denoted by $Sponge[f, pad, r]$. The sponge function takes two inputs: a bit string denoted by $N$ and the bit length denoted by $d$ of the output string denoted by $Z$. Note that the input $d$ determines the number of bits that the Sponge algorithm returns. But it does not affect the actual values. In principle, the output can be regarded as an infinite string whose computation is halted after the desired number of output bits is produced.

\begin{equation}
pad10 \cdot 1 (x,m) = 1||0^{(-m-1)mod(x)}||1
\end{equation}

The padding rule for $Keccak$ family is named multi-rate padding. Given a positive integer $x$, a non-negative integer $m$, the specification of padding rule for $Keccak$ denoted by $pad10 \cdot 1 (x,m)$ is specified as given in equation (11).

\subsection{SHA-3}

Keccak is the family of the sponge functions with $Keccak-p[b,12+2l]$ permutation. The family is parameterized by any choices of the rare $r$ and the capacity $c$ such that $r+c$ is in ${25,50,100,200,400,800,1600}$. When restricted to the case $b=1600$, the $Keccak$ family is denoted by $Keccak[c]$. In this case, $r$ is determined by the choice of $c$. The algorithm $Keccak[c]$ is specified as follows.

\begin{equation}
	Keccak[c](N,d)=Sponge[Keccak-p[1600, 24], pad10 \cdot 1, 1600 -c](N,d)
\end{equation}

SHA-3 hash functions and two SHA-3 $XOF$s will be defined. Given a message $M$, the four SHA-3 hash functions are defined from $Keccak[c](N,d)$ function by spending a two-bit suffix to $M$ and by specifying the length of the output as follows.

\begin{equation}
	SHA-3-224(M)=Keccak[448](M||01,224)
\end{equation}

\begin{equation}
SHA-3-256(M)=Keccak[512](M||01,256)
\end{equation}

\begin{equation}
SHA-3-384(M)=Keccak[768](M||01,384)
\end{equation}

\begin{equation}
SHA-3-512(M)=Keccak[1024](M||01,512)
\end{equation}

\begin{equation}
SHAKE-128(M,d)=Keccak[256](M||1111,d)
\end{equation}

\begin{equation}
SHAKE-256(M,d)=Keccak[512](M||1111,d)
\end{equation}

In this case, the capacity is double the digest length, in other words, $c=2d$ and the resulting input $N$ to $Keccak[c]$ is the message with the suffix appended $N=M||01$. The suffix supports domain separation, which distinguishes the inputs to $Keccak[c](N,d)$ arising from the SHA-3 hash functions from the inputs arising from the SHA-3 $XOF$s.

\section{GPU Accelerated SHA-3}

The SHA-3 parallel hash mode can be divided into two types: Batch mode and Tree mode. Batch mode is to divide a piece of information into multiple identical slices, and then hash the slices in parallel; or, Batch mode is to process multiple pieces of the same information in parallel at one time. Tree mode is to process multiple information into a form of Hash Root. In other words, multiple information is hashed in twos until the Hash Root is finally obtained. This Hash Root can be understood as Merkle Tree Root. In this article, we will adopt the Batch mode to implement the parallel calculation of the hash algorithm.

\subsection{Parallel Granularity}

SHA-3 parallel mode can be divided into Batch mode and Tree mode. The GPU parallelism used by different modes is different. For example, Batch mode normally uses 'one thread one message' parallel granularity which means that multiple GPU threads process multiple messages at the same time. Tree mode normally uses 'one thread one tree' parallel granularity which means that multiple GPU threads process multiple hash trees at the same time. In this paper, we mainly focus on Batch mode and our parallel granularity is 'one thread per message'.

\subsection{RC Tables Allocation}

SHA-3 RC tables are essentially look-up tables through which users can quickly implement part of cryptographic operations. Every thread needs to get access RC tables in each round of cryptographic operations. Hence we need to load RC tables in the memory of GPU in advance. In our benchmarking approach, we load RC tables into CUDA constant memory. Another possible solution is to load RC tables into CUDA share memory.

\subsection{Plaintext Allocation}

In our implementation of SHA-3 algorithm, we mainly use Batch mode. A large amount of messages are hashed at the same time. In our experiment, we hash a large amount of same length messages (10 bytes) via a large number of synchronized GPU threads. For example, if we have 100 messages and the length of each message is 10 bytes, then we need at least 100 CUDA threads for hashing. 

%Of course, if the scale of parallelism is larger, the performance of GPU can be further exerted.

\section{Experimental Results}

\begin{table}[]
	\centering
	\caption{Environment specification}
	{\footnotesize \label{my-label}
		\begin{tabular}{l|c}
			\hline
			CPU              & Intel Core i5-7200U @ 2.5GHz*4 \\ \hline
			GPU              & GeForce 940MX/PCIe/SSE2        \\ \hline
			Memory           & 12 GiB                         \\ \hline
			OS               & Ubuntu 16.04 LTS, 64 bits      \\ \hline
			CUDA compliation & V7.5.17                        \\ \hline
			GCC              & V5.4.0                         \\ \hline
	\end{tabular}}
\end{table}

Table 2 shows the configuration of our experiment platform. We conduct our experiments on a Ubuntu 16.04 LTS 64-bit operating system with a 12-GiB memory, an Intel Core i5-7200U @ 2.5GHz*4, an GeForce 940MX/PCIe/SSE2 GPU. The version of nvcc compliation is V7.5.17 and the version of GCC is V5.4.0. Our CPU code is written in standard C language and the GPU code is written in CUDA.

%Here we need to do some experimental results and analyze it.Here we need to do some imental results and analyze it.Here we need to do some experimental results and analyze it.

%Table 4 shows the performance between CPU based AES encryption and GPU based AES encryption. When the file size is small, say 1024 bytes, CPU performs better than GPU; however, when the file size is getting larger, say 10 KBytes, GPU performs much better than CPU. The performance of CPU is relatively stable while the performance of GPU can be highly excavated, meaning that GPU performance can be better used when more concurrent tasks are available.
% \usepackage{multirow}
\begin{table}[]
	\centering
	\caption{SHA-3 CPU V.S. GPU}
	{\footnotesize \label{my-label}
	\begin{tabular}{l|l|l|l|l}
		\hline
		\multirow{3}{*}{File Size (bytes)} & \multicolumn{4}{c}{Hash}                                                                                \\ \cline{2-5} 
		& \multicolumn{2}{c|}{Time (seconds)}                 & \multicolumn{2}{c}{Throughput (bytes per second)} \\ \cline{2-5} 
		& \multicolumn{1}{c|}{CPU} & \multicolumn{1}{c|}{GPU} & \multicolumn{1}{c|}{CPU} & \multicolumn{1}{c}{GPU} \\ \hline
		1202                               & 0.002656                 & 0.000431                 & 452560.24                & 2788863.11              \\ \hline
		4652                               & 0.008600                 & 0.000330                 & 540930.23                & 14096969.69             \\ \hline
		9302                               & 0.016400                 & 0.000330                 & 567195.12                & 28187878.78             \\ \hline
		18602                              & 0.032475                 & 0.000346                 & 572809.85                & 53763005.78             \\ \hline
		37202                              & 0.065230                 & 0.000373                 & 570320.40                & 99737265.42             \\ \hline
		74402                              & 0.129495                 & 0.000382                 & 574555.00                & 194769633.41            \\ \hline
		148802                             & 0.258204                 & 0.000437                 & 576296.26                & 340508009.15            \\ \hline
		297602                             & 0.516079                 & 0.000557                 & 576659.77                & 534294434.47            \\ \hline
		595202                             & 1.036339                 & 0.000755                 & 574331.37                & 788347019.87            \\ \hline
		1190402                            & 2.060367                 & 0.001204                 & 577762.12                & 988705980.06            \\ \hline
	\end{tabular}}
\end{table}

Table 3 shows the SHA-3 hashing performance comparison between CPU and GPU. One thing to note is that the GPU's parallel computing performance is affected by several factors, such as the ability of the GPU to perform parallel computing when the amount of concurrent tasks is small, such as where the file size is less than 1K bytes. Can't be played well. Once the number of parallelizable tasks is large, the GPU's parallel computing capabilities will be greatly utilized. In our experiments, it is very obvious that the performance of the GPU when the GPU performance is more than 1K bytes when the parallel file size is larger than the GPU performance. Exceeds the performance of the CPU by more than 4 times, and, as the number of parallelizable files increases, the GPU's powerful parallelism gets better.

\section{Conclusion}

This paper implements and optimizes Batch mode based Keccak algorithms on
NVIDIA GPU platform. Our work consider the case of processing multiple hash tasks at once and implement the case on CPU and GPU respectively. Our experimental results show that GPU performance is significantly higher than CPU is the case of processing large batches of small hash tasks. In future work, we aim to implement and analyze Hash Tree Mode based Keccak algorithms where many CUDA Reduce operations are involved. And our projected is now available at Github: https://github.com/Canhui/SHA3-ON-GPU.

%An important part of our work is optimizing the Hash Tree Mode based Keccak algorithms on
%NVIDIA GPU platform. In our work, the process of Tree Hashing is viewed as a group of CUDA
%Reduce operations. Moreover, several optimal strategies of Reduction are used to improve the
%performance of our algorithm. Finally, the algorithm shows that…

%Here we need to give out our conclusion.Here we need to give out our conclusion.Here we need to give out our conclusion.Here we need to give out our conclusion.Here we need to give out our conclusion.Here we need to give out our conclusion.Here we need to give out our conclusion.Here we need to give out our conclusion.Here we need to give out our conclusion.Here we need to give out our conclusion.

\section*{Acknowledgements}
This work is supported by Shenzhen Basic Research Grant SCI-2015-SZTIC-002.

{\small \bibliography{scibib}}

\bibliographystyle{ieeetran}

\end{document}